# Quantifying the search for solid Li-ion electrolyte materials by anion: a data-driven perspective

Austin D. Sendek,[1] Gowoon Cheon,[2] Mauro Pasta,[3] Evan J. Reed[1][†]

1. Department of Materials Science and Engineering, Stanford University, Stanford, CA, USA
2. Department of Applied Physics, Stanford University, Stanford, CA, USA
3. Department of Materials, Oxford University, Oxford, UK

[†]Corresponding author: evanreed@stanford.edu

## Abstract

We compile data and machine learned models of solid Li-ion electrolyte performance to assess the state of materials discovery efforts and build new insights for future efforts. Candidate electrolyte materials must satisfy several requirements, chief among them fast ionic conductivity and robust electrochemical stability. Considering these two requirements, we find new evidence to suggest that optimization of the sulfides for fast ionic conductivity and wide electrochemical stability may be more likely than optimization of the oxides, and that the oft-overlooked chlorides and bromides may be particularly promising families for Li-ion electrolytes. We also find that the nitrides and phosphides appear to be the most promising material families for electrolytes stable against Li-metal anodes. Furthermore, the spread of the existing data in performance space suggests that fast conducting materials that are stable against both Li metal and a >4V cathode are exceedingly rare, and that a multiple-electrolyte architecture is a more likely path to successfully realizing a solid-state Li metal battery by approximately an order of magnitude or more. Our model is validated by its reproduction of well-known trends that have emerged from the limited existing data in recent years, namely that the electronegativity of the lattice anion correlates with ionic conductivity and electrochemical stability. In this work, we leverage the existing data to make solid electrolyte performance trends quantitative for the first time, building a roadmap to complement material discovery efforts around desired material performance.

**I. Introduction**

As the world continues to strive toward a carbon-neutral economy to mitigate the effects of global climate change,[1] the need has never been greater for high performance battery technology that can reliably supply clean electricity to everything from electronic devices to automobiles to the grid. The realization of an all-solid-state lithium metal battery would be a tremendous development towards improving the energy density, power density, long-term stability and safety of energy storage devices. Janek and Zeir recently identified challenges that must be addressed in order to successfully develop SSBs and we refer to their paper for a comprehensive analysis.[2]

The sole replacement of liquid electrolyte with a solid leaves energy density unchanged and actually results in a decreased specific energy, as solids are denser than liquids. However, using high voltage cathodes (5V) would result in an energy density increase of 20%, while replacing graphite with Li metal could potentially increase it up to 70% due to the high specific capacity of Li metal.[2]

The principal technological bottleneck towards the goal of high energy density SSBs is identifying suitable solid electrolyte (SE) materials. SEs can be divided in two groups: inorganic solids (crystalline, glass or glass-ceramics) and organic solid polymers. The latter have the great engineering advantage of accommodating volume changes of the electrodes by plastic and elastic deformation and are more easily processable. Nonetheless, at room temperature their Li-ion conductivity is too low and their mechanical strength cannot prevent lithium dendrite growth, though improvements are being made on these fronts.[3,4] Among the inorganic SEs, crystalline materials offer the highest Li-ion conductivites, are generally considered the most promising candidates, and are therefore the focus of this work.

The search for crystalline solids with liquid-scale lithium conduction goes back into at least the 1970s,[5–9] but has accelerated considerably in recent years.[10–14] The major focus of this scientific undertaking has historically been on ionic conductivity, although with time it has become apparent that the search must involve more than simply looking for superionic behavior. For example, this realization tempered initial excitement surrounding reported superionic Li-ion conductivity in $Li_3P$ in the late 1980s[15] once it was discovered to have an unacceptably small electrochemical stability window. Recently, it has begun to appear that simultaneous optimization of ionic conductivity and electrochemical stability is a persistent difficulty. Today, the question remains: how to design materials with fast Li-ion conductivity that are stable against both Li metal and high voltage (>4V) cathodes? Many materials have been studied for both single- and multiple-electrolyte architectures, and this question continues to drive significant research efforts in both academia and industry. The optimization of select known materials (e.g. LiPON, LLZO) is a major part of this effort, as is the search for new materials. In this work we focus on materials discovery.

Enabling a rapid pace of materials discovery for solid-state batteries is imperative in light of the urgency of addressing the impending impacts of climate change and the considerable amount of time required to explore and characterize new electrolyte materials. It is critical to extract all information from the limited existing data in order to optimally inform future efforts. In this light, even a small amount of data has potential to go a long way.

Over the last several years, an important trend has emerged:[16] oxide-based Li-ion conductors generally tend to have good electrochemical stability but poor ionic conductivity, while sulphide-based Li-ion conductors have good ionic conductivity but poor electrochemical stability, although there are some important outliers to the trend. The electronegativity of the anion plays a direct role in setting the oxidation and reduction potentials and likely plays an important role in determining the energetic landscape in the material for ion conduction as well. This trend appears to extend to alternative anions, e.g. the fluorides, which have even higher stability and lower Li-ion conductivity than the oxides. While the anion has some predictive power for electrolyte performance, design principles beyond the anion identity have been reported recently,[17–20] suggesting additional contributing factors may be important. Recently it has also been proposed that fast conductivity and poor stability (and vice-versa) may be directly linked via lattice dynamics.[21,22]

Although the trends found in the existing data provide valuable guidance to searches for promising materials, these trends are based off of a small number of observations and they cannot yet *quantitatively* answer important broad questions: for example, how common are the high performance outliers? How much more (or less) likely are we to discover a fast conducting, highly stable oxide material than a sulfide material? Halide atoms possess electronegativities between oxygen and sulfur; are they more likely to simultaneously satisfy both the conduction and stability criteria? Is the multiple-electrolyte architecture more likely to succeed than the single-electrolyte architecture? Answering these questions quantitatively requires a bird's-eye view of the field informed by performance data.

Here we seek answers to these questions and more by aggregating and analysing the limited existing performance data in the literature and employing state-of-the-art predictive tools. We leverage the public materials database of the Materials Project[23] and its associated toolkit, and we incorporate our previous work in data-driven property prediction. The goal of this work is not necessarily to probe the physics of SE materials, but rather to quantitatively assess the state of the field through the lens of the existing data. This data sums up "what we know" thus far, serving as a Bayesian prior on our knowledge of materials space that enables us to quantitatively answer where the existing data suggests we should we look for promising new materials. We probe the likelihood of discovering superionic conducting materials with an electrochemical stability window width wider than 4V; superionic conducting materials with anodic stability beyond 4V vs. Li$^+$/Li, superionic conducting materials that are stable against lithium metal, and crystalline materials with a raw materials cost below the cost target of $10/m^2$. We also discuss the impact of mechanical properties and layered structure on ionic conductivity.

This effort to gain quantitative perspective from the whole of the available data might be interpreted as "the best we can do with the limited data we have," answering questions about the statistically expected performance of materials in a manner that is difficult with experimental or first principles methods. Importantly, we note that these trends are subject to change as more data on solid electrolytes is gathered, perhaps significantly so. We emphasize that this roadmap is only the first iteration of such a framework, and we welcome such changes over time. The accuracy and clarity of the trends in any data-driven roadmap will only improve in the future with each new data point. Due to the urgency of addressing our energy and climate challenges, it is imperative to seek new insights with the existing knowledge and tools, however imperfect they may be.

## II. Ionic conductivity and electrochemical stability

A SE material must exhibit high ionic conductivity and a wide electrochemical stability window, among additional requirements. In order to drive improvements beyond existing lithium-ion battery technology, this SE material must enable a battery voltage larger than 4V and ideally is stable against a Li metal anode. In considering the physical picture behind this simultaneous optimization, a difficult design challenge emerges. High ionic conductivity means that cation diffusion is extremely facile, while high electrochemical stability requires that diffusion of non-cation atoms is extremely difficult. One can envision a structure where both these criteria are simultaneously satisfied: the mobile cations move easily through a rigid sublattice that is stable with respect to the addition or removal of lithium. Although such an ideal structure may be conceptually simple, in reality these sharply differing electronic characteristics may not decouple completely from one another, resulting in higher electrostatic diffusion barriers and/or weaker sublattice bonds.

These potentially conflicting demands of solid electrolyte materials beg the question: given what has been discovered to date, can we quantify the likelihood of discovering a structure with both liquid-scale ionic conductivity and robust electrochemical stability? We probe this question of likelihoods for the first time with a data-driven approach.

Examining the interplay between ionic conductivity and electrochemical stability requires data on these properties for an overlapping set of materials. This overlapping dataset currently does not exist and therefore requires construction. For data on the electrochemical stability window, we leverage the density functional theory (DFT)-computed formation energies and band gaps of 6,600 materials as provided by the Materials Project.[23] In general, the band gap (in eV) provides an upper bound on the (thermodynamic) electrochemical stability window width (in V).[24,25] Because DFT tends to underestimate the true band gap,[26] we examined the DFT-computed band gaps against computed electrochemical stability window width data when available (see Section II, ii) and confirmed that the DFT-computed band gaps are, in most cases, effective upper bounds on the window widths. In this work, we use the DFT-computed band gap to predict an upper bound on the thermodynamic stability window width. To compute the (thermodynamic) oxidation and reduction potentials of each candidate, we draw upon the methods developed by Ceder et al. that leverage the DFT-computed grand potentials of various phases in the Materials Project database.[27–29] The oxidation and reduction potentials used in this work are those provided in the Materials Project. These potentials are only computed for materials that are thermodynamically stable (i.e. on the convex hull), limiting the number of candidate materials with predicted oxidation and reduction potentials to 723.

We compile electrochemical stability information for lithium-containing materials in the Materials Project database with the general formula Li-(X)-Y, where Y is one of the anions F, O, S, Cl, Br, P, I, N, Ge, or Si, and X is any combination of elements (except other anions). For the oxidation and reduction potential calculations there is no information on the germanides or silicides. For the band gap calculations we only include those materials that have a formation energy less than 0.1 eV/atom above the convex hull and are therefore predicted to be metastable and potentially synthesizable. These methods neglect potentially important kinetic factors for interfacial chemistry, and the accuracy of the oxidation/reduction potential calculations are dependent upon the completeness of the phase diagrams in the Materials Project, but we consider them to be the most accurate methods currently available. All structural data is downloaded from the Materials Project database through the REST API[30] and processed with the *pymatgen* Python library.[31]

Evaluating ionic conductivity in candidate materials requires either time-consuming experimental synthesis efforts, or computationally expensive methods like molecular dynamics simulations or diffusion barrier calculations for single crystals. It is likely that either approach requires at least several days to weeks per material, and computational approaches are unable to study the polycrystalline forms employed in devices. Instead of computing ionic conductivity in these many hundreds of materials, we employ a structure matching algorithm developed in our previous work.[18] This model compares essential structural features of candidate materials to those observed in 40 well-known, diverse Li-ion conductors and returns a percentage match between the candidate material and the fast conducting reference materials (the "superionic likelihood"). These features include the lithium-lithium bond number (LLB), the anion framework coordination (AFC), the sublattice bond ionicity (SBI), the Li-anion separation distance (LASD) and the Li-Li separation distance (LLSD).

The accuracy of this structure-matching model has been extensively benchmarked against DFT simulations,[32] and is reported to identify room temperature superionic Li conductors approximately three times more effectively than completely random guesswork. In this previous benchmarking study, the F1 score for correct superionic identification of the model is reported to be 0.5. The F1 score for correct superionic identification via completely random guesswork is reported to be 0.14. This suggests the model possesses a useful degree of predictive power. Furthermore, the DFT evidence suggests the materials identified by this model possess a log-average conductivity more than 44 times greater than that of randomly chosen materials. Importantly, this model is at least 5 orders of magnitude faster to evaluate than density functional theory simulations. Additionally, the model is trained on experimental conductivity measurements where microstructure may play a role; DFT simulations are limited to the range of hundreds of atoms, and so such predictions are beyond the capability of DFT and a data-driven approach is required. To our knowledge, there exists no other benchmarked model for rapidly predicting superionic conductivity in chemically diverse structures. Although we do not expect 100% accuracy from the current form of this model, it does capture general structural trends based on the existing data used for training. For the purposes of this analysis, we assume the model represents the "ground truth" of conductivity. We expect this model to improve as the amount of data and collective wisdom surrounding Li ion conductors grows.

The form of the structure-matching model is the following: $P_{superionic} = [1 + \exp(-\theta^T x)]^{-1}$, where:

$$\theta^T x = 0.184 \times LLB - 4.009 \times SBI - 0.467 \times AFC + 8.699 \times LASD - 2.170 \times LLSD - 6.564$$

We refer the reader to Section 3.3 and the Supplemental Information of Ref. [18] for further discussion of these structural features and their implication for design thinking in superionic conductors.

### i.     *Optimizing for a wide stability window*

Using the band gap as an upper bound on the electrochemical window width, we plot this value against the superionic likelihood for all 6,600 materials in Fig. 1(a). We note that DFT formation energy calculations predict only 11% (723) of these 6,600 candidates to be on the convex hull, i.e. are thermodynamically stable. The remaining 89% of these materials may be metastable. Training a model on as large a data set as possible is crucial for any data-driven effort, and we justify using the band gap instead of the DFT-computed thermodynamic electrochemical stability window width because it allows us to train on over eight times more data points. We leverage the DFT-based grand potential methods in the following subsections *ii*, *iii*, and *iv* where specific potential values are desired rather than window widths.

In total, we compute performance prediction data for 5,350 oxides, 790 fluorides, 172 sulfides, 21 bromides, 46 chlorides, 11 iodides, 75 nitrides, 33 phosphides, 62 germanides, and 40 silicides. Although there is significant variance among the data in each family, figure 1(a) reproduces the performance trends known among the community: the sulphides generally conduct well but have poor stability; the oxides have poor conductivity and high stability. The fluorides have low conductivity and high stability as well. The alignment of these trends with the conventional wisdom builds confidence in our modeling approach. Relatively few materials are in the promising upper-right quadrant of Figure 1(a), reinforcing the difficulty of simultaneous optimization.

We replace the data for each family with its centroid (mean value) in the two-dimensional performance space in Fig. 1(b) to clarify the performance distribution of each family. An ellipse that delineates one standard deviation of the two-dimensional data is plotted to guide to the eye. Although the data is fairly noisy, there are clear cases where the distance between means of families is smaller than the standard deviations (e.g. the oxides versus the sulphides). This plot also shows the approximate trend that the average structural similarity to known ion conductors (predicted superionic likelihood) decreases and the electrochemical stability increases as the electronegativity of the anion increases. The strongly electronegative anion families are more likely to form strong ionic bonds, apparently improving stability but decreasing conductivity. Pure lithium metal represents the limit of zero electronegativity difference between Li atoms and the anion lattice, yielding a notoriously reactive but superb lithium conductor.

We have also extended this analysis to the polyanionic oxide compounds, where the anion Y may be one of $P_xO_y$, $B_xO_y$, $Si_xO_y$, $F_xO_y$, $C_xO_y$, $Sb_xO_y$, $Sn_xO_y$, $Te_xO_y$, $S_xO_y$, $Ge_xO_y$, $N_xO_y$, or $Cl_xO_y$. We find the general performance trends tend to mirror that of the oxides, and due to this similarity we exclude this data from Fig 1.

To quantify these trends we interpret the two-dimensional data distributions for each family plotted in Fig. 1(a) as a Bayesian prior on electrolyte performance as a function of anion. If future research in each family were to blindly sample from the same probability distribution as these currently known materials, these distributions represent the expected performance of the new materials as a function of anion choice. In order to calculate the likelihood of encountering a high performance outlier through random sampling in each family $P_{\text{outlier}}$, we find the fraction of the underlying data distributions that lie beyond the minimum required metrics of 50% likelihood of superionic conduction and a minimum window width of 4V. We consider these materials to be the most promising candidates. This analysis allows for direct and quantitative comparison of "promise" between different anion families as interpreted by these models.

We note that the current state of research does not necessarily follow these conditions, i.e. materials are not chosen at random based only on the anion. In practice, researchers may only study a subset of a given anion family and may employ design principles with more predictive power than random guesswork. Indeed, some materials discovery efforts are more likely to succeed than others depending on the strategy employed. We emphasize that $P_{outlier}$ does not strictly quantify the likelihood of a breakthrough discovery coming from any given research effort, but rather quantifies the broad performance of entire material families based only on the anion. Rather than classifying the families by anion, one could classify materials by structural motifs e.g. garnets, which could give different values of $P_{outlier}$.

The underlying distributions are not well-known, and are difficult to infer with small amounts of data. We note in some cases the data approximately follows Gaussian behavior, e.g. the oxides in Fig. 1(c), while sometimes the data is less Gaussian, e.g. the bromides in Fig. 1(a). This behavior is also difficult to observe by eye because much of the data in Figs. 1(a), (c), and (e) is obscured by other data points; for context, there are more than 5,000 oxide data points plotted in Fig. 1(a) with most data clustered tightly around the mean.

Without clear knowledge of the underlying distributions, we compute the probability of discovering promising new materials from each family by counting the fraction of materials from each family above the minimum required performance metrics ("binning"). It is unavoidable that not every promising candidate material will demonstrate superior performance experimentally (i.e. the models have some error), but we expect the overall performance distribution of each family to possess a degree of statistical accuracy.

The performance of material $i$ is defined by its superionic likelihood $P_{\text{superionic}}^{(i)}$ and predicted electrochemical stability window width $V_{\text{w}}^{(i)}$. The binning approach counts the fraction of materials in each family with $P_{\text{superionic}} > 50\%$ and $V_{\text{w}} > 4\text{V}$:

$$P_{\text{outlier}} = \frac{\sum_{i=1}^{N} \mathbb{I}(P_{\text{superionic}}^{(i)} > 50\%) \times \mathbb{I}(V_{\text{w}}^{(i)} > 4\text{ V})}{N}$$

Here the $\mathbb{I}(X)$ symbol denotes the indicator function, which evaluates to 1 if the argument $X$ is true and 0 otherwise. The numerator represents the number of materials in the upper-right quadrant of the graph for a given family, and the denominator represents the total number of materials in that family. The likelihoods for each family are provided in the second-to-left column of Table 1.

In addition to this binning method, we also compute the likelihood of outliers in each family by fitting the data to a two-dimensional Gaussian distribution and then finding the cumulative probability of the distribution beyond the minimum required performance metrics via integration. This method relies on the data being approximately normally distributed. This method is detailed in the Supplemental Information, Section S1 along with an assessment of the Gaussian character of the data via the Shapiro-Wilk test for normality[33] in Figure S1. We provide the results of this alternative test in parentheses in Table 1. In most cases the results from the two methods align within approximately a factor of two.

These data predict that the likelihood of discovery of a fast Li conducting material with a stability window width beyond 4V from all known oxides, given the existing data, is quantitatively much smaller than other anion families. This might suggest that continued efforts in searching the oxides generally for SEs in single-electrolyte architectures are better invested in other families, although there has been interest in oxides for other applications including resistive electronic memories. The fluorides are also a poor prospect for this battery architecture. The Gaussian integration method suggests the likelihood of high performance outliers coming from the sulphides is relatively low, yet higher than for the oxides and fluorides. From this data, we derive the following insight: given what has been discovered so far, one is more likely to find a fast conducting, >4 V window sulphide outlier than an oxide. Essentially, this suggests one is marginally more likely to overcome stability issues with sulphides than conductivity issues with oxides.

"Uneven sampling of these distributions has the potential to impact uncertainty in the computed $P_{outlier}$. The poor performance of the oxides generally is unavoidably affected by the fact that a large number of Li-containing oxides are known, having been synthesized and characterized in much higher volume many years before the other families. By contrast, Li-containing sulphides have only been the focus of major research efforts within the past decade, making the average known Li-containing sulphide a much better conductor than the average known Li-containing oxide. The conclusions claimed here pertain to random selection from the family as a whole based on the distribution of known materials, and should not be interpreted as commenting on the promise of specific structure types within each family, e.g. the garnet oxides."

Both methods reveal that the most promising classes appear to be the often overlooked halides: the chlorides, bromides, and iodides show the highest likelihoods. These materials have been less exhaustively investigated than the oxides and sulphides, so a much smaller pool of data is available for both model training and for screening (for screening, there are only 46 chlorides, 21 bromides, and 11 iodides, versus 5350 oxides), but the existing data suggest new efforts in this area may be worth the investment. The phosphides, silicides, and germanides appear to have major stability issues and thus may not be appropriate for battery applications, although their average conduction properties are much better than other families.

For comparison, we also calculate the likelihood of discovery of materials only with window width above 4 V, and not considering our superionic structure matching model. These results are provided in Table 1. With this selection criteria, the halides remain promising and the promise of the (poorly conducting) fluorides increases, as we might expect.

It should be reiterated that there is significant variation within each family, and the likelihood data in Table 1 is a statistical average over the performance data of all known materials in each family. Taking additional design principles into consideration during the materials search beyond just anion choice may result in better observed material performance on average. Furthermore, we note that, in practice, such strong constraints on electrochemical stability may not be necessary, as the solid-electrolyte interphase (SEI) may stabilize the electrolyte against interfacial reactions.[34] However, without *a priori* knowledge of the interfacial chemistry arising from a given electrolyte/electrode combination, we assume these strict demands must be met in candidate materials in order to minimize false positives.

*ii. Stability against high voltage cathodes*

An alternative strategy for enabling high energy density solid-state batteries (SSBs) is to use a multiple-electrolyte architecture, where a high oxidation potential electrolyte/coating is placed on the cathode and a low reduction potential electrolyte/coating is placed on the anode as proposed by Richards et al.[29] In this approach, the objective of a materials search for cathode coatings is to discover a fast Li-ion conductor material with an oxidation potential beyond 4 V vs. $Li^+$/Li. Ideally, the cathode voltages in candidate SSB chemistries should be 5V vs. $Li^+$/Li or higher to drive significant improvement over existing LIB technology.

We compute the superionic likelihood and predicted oxidation potential for as many candidates as possible and quantify the likelihood of discovering outliers in each family. Here we consider the oxidation potentials listed in the Materials Project database for 723 materials based on DFT grand potential methods. In all, we consider 427 oxides, 56 sulfides, 97 fluorides, 25 chlorides, 7 bromides, 6 iodides, 73 nitrides, and 32 phosphides. With this smaller pool of materials under consideration, there is no data available for the silicides or germanides.

In Fig. 1(c) we plot the predicted oxidation potential versus predicted superionic likelihood for all 723 entries, and in Fig. 1(d) we replace the data with the centroids and standard deviations to guide the eye. In Fig. 1(d), we see a moderate inverse correlation (Pearson correlation coefficient of -0.5) between ionic conductivity and oxidation potential. In this case the trend of ionic conductivity with oxidation potential is more pronounced than it was with ionic conductivity and stability window width in Fig 1(b) (which had only a Pearson correlation coefficient of -0.1). The electronegativities of the anions are provided in Fig. 1(d) for reference. The oxides and sulfides appear to be too ionically insulating and too unstable, respectively. We note again that the halides, with their electronegativities somewhere between the too-stable oxides and the too-unstable sulphides, appear to be particularly promising candidates for this purpose.

We bin this data as before, now using the bounds of >50% superionic likelihood and oxidation potential $V_{ox} \geq 4V$:

$$P_{\text{outlier}} = \frac{\sum_{i=1}^{N} \mathbb{1}(P_{\text{superionic}}^{(i)} > 50\%) \times \mathbb{1}(V_{ox}^{(i)} \geq 4 \text{ V})}{N}$$

These likelihoods are provided in Table 1, along with the likelihoods from the Gaussian method.

Cathode coatings may be made thin enough that materials with poor ionic conductivity are acceptable. In this case, a materials search would only seek high anodic stability. We compute the likelihood of finding materials with high oxidation potential, regardless of predicted ionic conductivity, for each family and provide the results in Table 1. Perhaps unsurprisingly, the fluorides are the strongest candidates by far. Supporting this conclusion, recent work has demonstrated that fluoride-based cathode coatings can minimize electrolyte degradation and improve cycle life if they can be made thin enough that ionic impedance is minimized.[35]

### iii. Stability against lithium metal

The multiple-electrolyte architecture also requires the discovery of fast Li-ion conducting materials that are stable against Li metal as the anode-facing electrolyte. Using the same pool of candidate materials and the same methods as in the previous section, we analyze the reduction potential of all 723 candidate materials against the superionic likelihood. In Fig. 1(e) we plot the results for all materials, and in Fig. 1(f) we visualize the distributions. The likelihoods for discovery in this case are found by counting those materials with a reduction potential of $V_{\text{red}} \leq 0$ V:

$$P_{\text{outlier}} = \frac{\sum_{i=1}^{N} \mathbb{1}(P_{\text{superionic}}^{(i)} > 50\%) \times \mathbb{1}(V_{\text{red}}^{(i)} \leq 0 \text{ V})}{N}$$

These results are provided in Table 1. We observe particularly high stability against Li metal in the nitrides, a trend recently noted by Zhu et al.[36] The phosphides also appear to be promising candidates. Li$_3$P is a case in point: its electrochemical window is narrow (~1V), but it is a fast ion conductor that is stable against Li metal.[29] It has been experimentally observed to form at the Li-metal/LGPS interface, where its fast ion conduction properties are particularly beneficial.[37]

As is the case with cathode coatings, anode coatings may be made thin enough that ionic conductivity becomes a less important constraint. In this case, we compute the likelihoods of outlier discovery of a material with high stability against Li metal in each family, regardless of predicted ionic conductivity, and provide them Table 1. This is the only category where all anion families considered have a nonzero likelihood, although the nitrides are much more likely than others.

iv. *Fast ionic conductivity, stability against Li metal, and high oxidation potential: does the "wonder material" exist?*

The ideal electrolyte material is a fast ion conductor that is stable against lithium metal and simultaneously has a high oxidation potential. Leveraging the known likelihoods for zero reduction potential and oxidation potential given in Table 1, we compute the joint probability that a material from a given anion family is stable against lithium metal and has an oxidation potential $V_{ox}$ that is above a minimum $V_{ox,min}$. Treating the oxidation potential as independent of the reduction potential, the intersection of these two requirements requires a product of probabilities:

$$P_{\text{outlier}}(V_{\text{red}} \leq 0, V_{\text{ox}} \geq V_{\text{ox,min}}) = P_{\text{outlier}}(V_{\text{red}} \leq 0) \times P_{\text{outlier}}(V_{\text{ox}} \geq V_{\text{ox,min}})$$

The results of this calculation are provided for $V_{ox,min}$ = 4 V Table 1, as a function of $V_{ox,min}$ in Fig. 2 using the Gaussian integration method. When ionic conductivity is considered, the phosphides and nitrides are the only families with significant likelihoods, although they fall to zero near $V_{ox,min}$ = 2V. The chlorides are the only family with a non-zero likelihood (0.6%) at $V_{ox,min}$ = 4V. This means one is very unlikely to discover an outlier material in any family with fast ion conduction, stability against lithium metal, and >4V stability against oxidation. In fact, according to the Gaussian method, this data suggests that superionic materials with a >4V stability window *and* stability against Li metal are approximately two orders of magnitude less common than superionic materials with only a >4V stability window (and arbitrary reduction potential). The outlook is more promising when ionic conductivity is not considered; here the fluorides, oxides, chlorides, and bromides have nonzero likelihood at $V_{ox,min}$ = 4V. This has implications for the design of solid-state Li-ion batteries with Li metal anodes: it suggests a dual electrolyte architecture is much more likely to be able to meet necessary material performance requirements.

*v. Discussion*

The results in Fig. 1 suggest some correlation exists between ionic conductivity and electrochemical stability. This negative trend is particularly apparent between oxidation potential and ionic conductivity, where the electronegativity of the anion in the lattice appears to play a significant role.

This negative correlation, and the relatively small discovery likelihood percentages in Table 1, quantitatively reinforce that simultaneously optimizing a material for both ionic conductivity and a wide electrochemical stability window is a difficult task indeed. It appears to be even more difficult when requiring stability against lithium metal. Judging by their relative positions in performance space, exploring oxide and sulphide materials appears to be more difficult than the halides. The odds are not much higher for the multiple-electrolyte strategy either, although the data does offer a most likely way forward: a nitride-based SE on the anode, and a chloride- or bromide-based SE on the cathode or a thin fluoride-based cathode coating. We note a thin fluoride-based cathode coating (~90% likelihood) and a fast conducting nitride-based anode coating (~10% likelihood) has similar predicted likelihood of success (90% × 10% = 9%) to the discovery of a fast conducting bromide with a window width beyond 4V (~10%). However, the likelihood of discovering a chloride "wonder material" that is stable against Li metal (~4%) and has an oxidation potential of >4V (~16%) is more than an order of magnitude less likely using the binning method (4%×16%=0.6%) and two orders of magnitude less likely using the Gaussian method (6%×1%=0.06%), suggesting that the successful realization of a SSBs with Li metal anode is more likely with a dual electrolyte architecture.

In the Supplemental Information Fig. S2 we provide the outlier discovery likelihoods computed with the Gaussian integration method for arbitrary window widths, minimum oxidation potentials, and maximum reduction potentials, with and without requiring fast ionic conductivity.

The small magnitude of the outlier discovery likelihood values computed in this section also highlight the importance of considering kinetic electrochemical stability in candidate materials. Although some material families have significant discovery likelihoods for certain applications, the data suggests the outlook for simultaneous conductivity/stability optimization is generally poor when requiring thermodynamic stability. However, kinetic stabilization may enable the use of fast Li-ion conductors with limited thermodynamic stability. To our knowledge, no data-driven or first principles tools exist for rapidly assessing the kinetic stability of materials exposed to a chemical potential, although such tools would clearly be of high value to the community's ability to perform efficient searches of materials space.

## III. Additional correlations with ionic conductivity: mechanical properties, layered structure

Mechanical properties are an important contributor to overall solid electrolyte performance. There are several complex and potentially conflicting mechanical demands put on candidate SE materials, e.g. a material should be "soft" enough to maintain contact with rough electrode surfaces while simultaneously being "hard" enough to block dendritic growth and withstand abuse without cracking. The development of fast, machine learned models for predicting elastic properties is an ongoing area of research,[38] and large amounts of elastic property data on Li-containing materials do not yet exist. This makes extrapolating from the small amount of existing materials data difficult.

Recent work from the Ong group[39] reported computed elastic properties for 23 well-known inorganic crystalline SE materials. Their report suggests the shear modulus (which was used as a metric for stability against dendritic growth) is correlated with the electronegativity of the anion in the lattice and is inversely correlated with ionic conductivity. If high ionic conductivity, high shear modulus SE materials are desired, an additional difficult optimization process may lie beyond the electrochemical stability/ionic conductivity optimization.

We find that structural layeredness is another property with inverse correlations to ionic conductivity. Today's lithium-ion batteries utilize layered materials to achieve lithium intercalation on both cathode and anode; superionic layered materials could feasibly function as solid-state electrolytes as well. Having identified the subset of layered structures that exhibit weak interlayer bonding and strong intra-layer bonding in the Materials Project database in our previous work[40], we compute the probability that a stable material is predicted to contain the necessary structural characteristcs to be a fast ion conductor, given that it is a layered structure with intercalated lithium, is only 5%. This is a factor of two lower than the predicted 10% background probability that a stable lithium containing solid is a superionic conductor. The posterior distribution is similar: the likelihood of being layered, given that a material is predicted to be a superionic conductor, is 10%; also a factor of two lower than the 21% background probability that any stable Li-containing compound is layered.

The existing data suggests that layeredness decreases the likelihood of fast ion conduction in a candidate material by half. Although undiscovered layered superionic conductors may exist, we predict layeredness itself is a poor indicator of fast ion conduction. This suggests that optimizing layered cathode materials for fast rate capability is likely to be more difficult than optimizing non-layered cathode materials, although bulk materials generally introduce an additional difficulty with electrochemical expansion upon lithiation.

## IV. Costs and manufacturing

The cost of solid electrolyte materials is a metric of critical importance. The U.S. Department of Energy's Advanced Research Projects Agency – Energy (ARPA-E) has adopted an aggressive cost target of $10 (USD)/m$^2$ for electrolyte materials, as originally proposed by McCloskey.[41] The overall cost is broken down into materials cost and processing cost.

In Fig. 3 we plot the median raw materials cost per area per 10 μm thickness for the materials in each family, following the values provided in the Supplemental Information of ref. [18]. Overlaid on this graph is the materials cost per mole of the anion. Raw materials costs change in time, and several of these reference values are already significantly different from current prices; in most cases the current prices are lower than utilized in Fig. 3. Even when recognizing these values are likely upper bounds on material costs, the median cost for all families except the fluorides and germanides is below the $10/m$^2$ threshold. The standard deviation in the cost data is very large. However with these cost values we find that 70% of all 6600 candidate materials considered here fall below this cost target before processing costs are considered. Since it is more likely than not that a candidate material will meet this cost target, we conclude materials cost is unlikely to be a significant barrier and the larger amount of material required for multiple-electrolyte architectures may be acceptable.

Excluding compositions with particularly expensive elements (e.g. Sc, Ge, Rb, Ru, Rh, Cs, Yb, Os, etc.) from consideration will increase the likelihood of candidate materials being under the cost target. Making electrolytes thinner will also scale materials costs down linearly, but manufacture of thin ceramic materials is difficult and manufacturing costs will likely begin to increase as the materials approach single micron thickness or thinner. In addition, many materials often become very brittle in this thickness regime; brittle electrolytes may crack or pulverize during use, decreasing ionic contact and enabling dendritic growth. In this respect, cost concerns would dictate that candidate SE materials should be made flexible or should have high fracture toughness.

The manufacturing processes and costs of materials are difficult to generalize, but the anion in the structure can have a significant effect on the methods and ease of synthesis at the laboratory scale. Oxides have perhaps the greatest ease of manufacture due to the natural oxidation that occurs when precursors are exposed to air. Most other compounds must be synthesized either in a dry room or under inert gas or vacuum in order to minimize unwanted reactions with atmospheric oxygen. Sulfides and phosphides, the two most electropositive anions considered here, have additional issues with water sensitivity and therefore must be carefully sealed to minimize unwanted hydrolysis. Fluorides are often synthesized from HF gas, which reacts with moisture in the air to form dangerous hydrofluoric acid. The halides, highlighted in the previous section for their promise, can typically be synthesized in solution, a much easier process than the solid state synthesis required of sulfides. This may make the halides an even more attractive family for laboratory scale study.

These various restrictions make laboratory scale synthesis of certain conductor families difficult, creating a barrier to studying candidates outside of the oxides. Moving outside the oxides, however, appears to be critical to driving the field forward. Exploring these alternative families will likely increase financial and time costs for research laboratories, but costs and other manufacturing barriers tend to scale away at the industrial level. The cost of synthesizing and studying Li conductors from unconventional families may be high in the short run, creating barriers to new materials discovery, but once these high performance materials are discovered the anion type will likely have limited to no impact on the ability for promising new chemistries to hit industrial cost targets.

## V. Conclusions

By taking a data-driven approach, we uncover correlations between several fundamental performance metrics in SEs that would be difficult to observe from first principles approaches. To summarize and quantify these correlations, we compute the Pearson correlation coefficients between ionic conductivity, electrochemical window width/band gap, oxidation potential, reduction potential, materials cost, and anion electronegativity for all candidate materials. These correlation coefficients are provided in Table 2. Although some metrics such as cost have low correlation with other metrics, we find the magnitude of the correlation coefficient is as high as 0.5 between ionic conductivity and oxidation potential.

We draw four central conclusions from this data: (1) some material families are more likely to yield high performance outliers for certain applications, and this is quantified in Table 1; (2) the coating/multiple-electrolyte strategy appears to have a similar likelihood of success to finding a >4V stability window material for a one-electrolyte architecture, although the multiple-electrolyte architecture is over 10 times more likely to be successful when stability against lithium is required, (3) the low magnitudes of outlier discovery likelihoods for thermodynamically electrochemically stable fast ion conductors in general highlight the importance of considering kinetic stabilization of fast ion conducting phases, and (4) these low magnitudes also underscore the importance of implementing targeted searches of material space informed by data and design principles.

On conclusion 1, we recommend new searches of the chlorides and bromides for fast Li-ion conducting, electrochemically stable materials. Although the volume of data in this space is limited, the existing data points suggest strong overall performance in these families. Recent reports suggest some promising new materials may be emerging from these families.[42] The nitrides appear to be an exciting class of materials for Li-ion conducting materials that are stable against Li-metal. The distribution of performance data in the oxides and the sulphides suggest they are less likely to produce the high performance materials required for the SSBs of the future. On conclusion 2, we note a single-electrolyte battery with stability against Li metal appears to be much less likely than a multiple-electrolyte Li metal battery as proposed by Richards et al.[29] This suggests that efforts to develop SSBs with Li metal anodes are more likely to succeed if a multiple-electrolyte architecture is employed. Most candidate materials are under the DOE cost target of $10/m$^2$ when considering raw materials costs only, so thicker electrolytes with multiple layers may be acceptable. The manufacturing costs of these architectures is likely to be higher, however, and this may ultimately determine the cost competitiveness of this strategy.

On conclusion 3, we note that complete thermodynamic stability is a fairly strict requirement for candidate materials, significantly decreasing the likelihood of discovery in most cases. However, some of these thermodynamically unstable materials may be kinetically stable; requiring thermodynamic stability may introduce some false negatives. Future searches should consider kinetic stability, and the development of new tools for rapid prediction of kinetic stability from atomistic structure would be of great value to the community. Finally, on conclusion 4, the existing data warns against random searches of materials space, particularly in the oxides and sulphides. Emerging design principles and data-driven efforts provide a new route for beating the odds and identifying the promising outliers. The addition of new data could change some of the conclusions presented here. However, it is clear that the continued development of data-driven predictive models is of utmost importance to seeing accelerated progress in the field, and we expect many exciting new advancements as these efforts expand.

## VI.     Acknowledgements

This work is supported in part by a seed grant from the TomKat Center for Sustainable Energy at Stanford University and in part by Toyota Research Institute through the Accelerated Materials Design and Discovery program. The authors thank Prof. Nicole Adelstein, Gowoon Cheon, Dr. Ekin D. Cubuk, William E. Gent, Leonid Kahle, Ruby A. Lai, Prof. Nian Liu, Prof. Robert Sinclair, Dr. Brandon Wood, and Qian Yang for constructive conversations on these ideas.

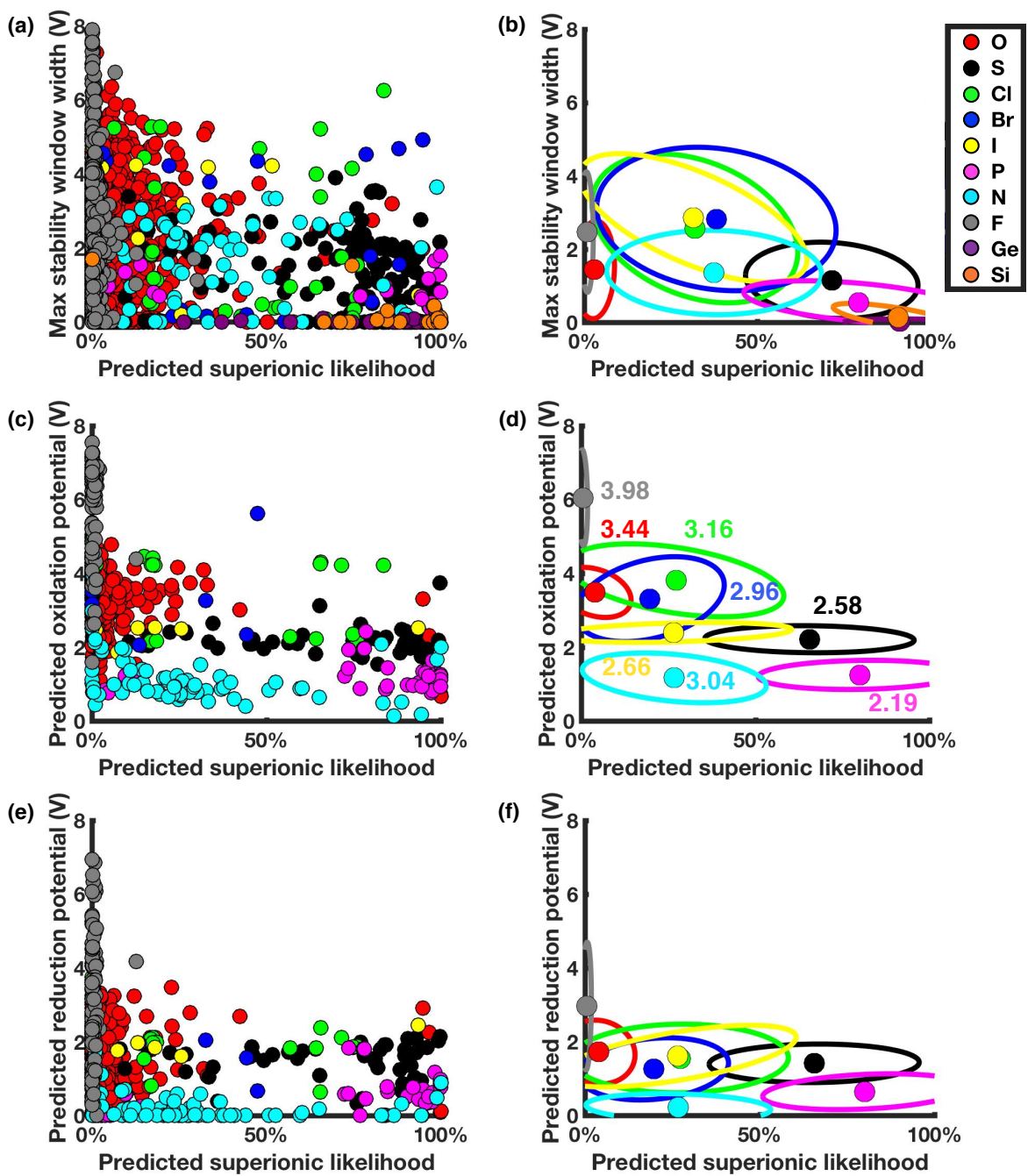

**Figure 1:** *The property spectrum of solid Li-ion conductor candidates by anion family.* We plot predicted metrics of electrochemical stability against predicted likelihood of superionic

conduction for stable Li-containing crystalline solid materials from the Materials Project database. These materials have the general composition Li-(X)-Y, where (X) is any combination of elements and Y is an anion from the following group: F, O, S, Cl, Br, I, N, P, F, Si, or Ge. In figure **(a)** we plot the predicted electrochemical window width (based on the DFT-calculated band gap in the Materials Project) against the superionic likelihood for the stable Li-containing materials. Materials in the upper right quadrant of the plot (>50% chance of fast ion conduction and >4V stability window width) are considered promising candidates. In figure **(b)** we plot the centroid of each family and the ellipse that captures one standard deviation of the data as a rough guide to the eye. The chlorides, bromides, and iodides appear to be promising candidates for wide stability window, fast ion conducting structures. In figures **(c)** and **(d)** we plot the predicted thermodynamic oxidation potential (based on the DFT-calculated formation energies and the phase diagram tools in the Materials Project) against the superionic likelihood and see that the fluorides are particularly robust against oxidation while the nitrides and phosphides are not. The anion electronegativity appears to be positively correlated with the oxidation potential and negatively correlated with superionic likelihood; the electronegativities of the anions are provided as reference. In figures **(e)** and **(f)** we plot the predicted reduction potential (from the Materials Project) against the superionic likelihood and note that the nitrides and phosphides are particularly stable against lithium.

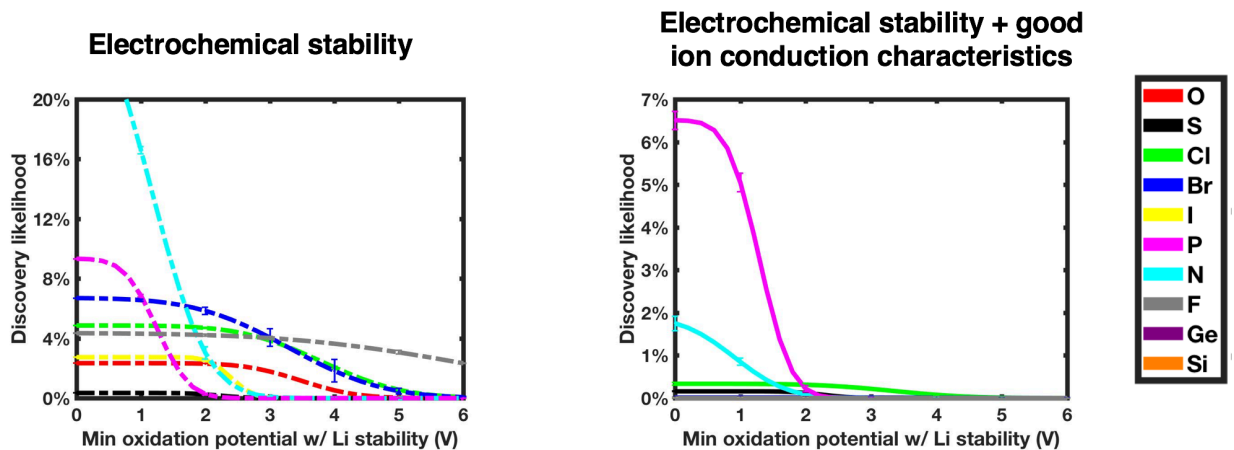

**Figure 2:** *Outlier discovery likelihood for superionic materials stable against oxidation at given voltage and reduction by Li metal.* In the left pane we use the Gaussian integration method to calculate the likelihood of discovering a material with stability against Li metal and an oxidation potential above the value on the x-axis. Here the requirement of fast Li-ion conductivity is ignored. The discovery likelihood at an oxidation potential of 4 V is highest (3.9%) for the fluorides and is also non-zero at this potential for the chlorides, bromides, and oxides. In the right pane, we add in the ionic conductivity constraint. Here the chlorides are the most promising family for fast Li-ion conduction, Li metal stability, and oxidation stability at 4 V, but the likelihood is several times lower than before. All other families have a discovery likelihood of essentially zero. From this, we conclude that the likelihood of finding two fast Li-ion conducting materials, one of which that is stable against reduction at 0 V and one of which is stable against oxidation at 4 V, is much more likely than finding a single material with these properties. This suggests two-electrolyte architectures are more likely to succeed for solid-state Li metal batteries than single-electrolyte architectures.

| | Electrochemical stability and ionic conductivity | | | | Electrochemical stability only | | | | Ionic conductivity only |
|---|---|---|---|---|---|---|---|---|---|
| Anion family | Likely fast Li-ion conductor, >4 V window width | Likely fast Li-ion conductor, stable above 4V | Likely fast Li-ion conductor, stable at 0V | Likely fast Li-ion conductor, stable at 0V and above 4V | Any Li-ion conductivity, >4V window width | Any Li-ion conductivity, stable above 4V | Any Li-ion conductivity, stable at 0V | Any Li-ion conductivity, stable at 0V and above 4V | Mean likelihood of fast Li-ion conduction, any stability |
| F | <0.1% (<0.1%) | <0.1% (<0.1%) | <0.1% (<0.1%) | <0.1% (<0.1%) | 9.1% (17%) | 91% (84%) | 1.0% (4.4%) | 0.9% (3.7%) | 0.8% |
| O | <0.1% (<0.1%) | <0.1% (<0.1%) | <0.1% (<0.1%) | <0.1% (<0.1%) | 3.6% (2.7%) | 18% (22%) | 2.8% (2.3%) | 0.5% (0.5%) | 3.0% |
| Cl | **15% (4.1%)** | **16% (6.3%)** | 4.0% (1.2%) | 0.6% (<0.1%) | **35% (24%)** | **68% (42%)** | 12% (4.9%) | 8.2% (2.1%) | 32% |
| N | <0.1% (0.4%) | <0.1% (<0.1%) | **9.6% (8.5%)** | <0.1% (<0.1%) | <0.1% (1%) | <0.1% (<0.1%) | **51% (28%)** | <0.1% (<0.1%) | 38% |
| Br | **14% (9.1%)** | <0.1% (4.0%) | <0.1% (0.2%) | <0.1% (<0.1%) | **52% (27%)** | 14% (27%) | 14% (6.7%) | 2.0% (1.8%) | 38% |
| I | **9.1% (1.3%)** | <0.1% (<0.1%) | <0.1% (<0.1%) | <0.1% (<0.1%) | **55% (26%)** | <0.1% (<0.1%) | 17% (2.7%) | <0.1% (<0.1%) | 32% |
| S | <0.1% (0.3%) | <0.1% (<0.1%) | <0.1% (0.2%) | <0.1% (<0.1%) | <0.1% (0.3%) | <0.1% (<0.1%) | 1.8% (0.3%) | <0.1% (<0.1%) | 72% |
| P | <0.1% (<0.1%) | <0.1% (<0.1%) | **9.4% (8.0%)** | <0.1% (<0.1%) | <0.1% (<0.1%) | <0.1% (<0.1%) | 16% (9.3%) | <0.1% (<0.1%) | 79% |
| Ge | <0.1% (<0.1%) | - | - | - | <0.1% (<0.1%) | - | - | - | 91% |
| Si | <0.1% (<0.1%) | - | - | - | <0.1% (<0.1%) | - | - | - | 91% |

*Likelihood of high performance outliers based on existing data*

**Table 1:** *Likelihood of new outlier discovery by anion family.* For each of the anion families, we calculate the percentage of the underlying data distribution that exists in the high performance region of performance space. This percentage represents the likelihood that a new stable material from this family will satisfy these requirements if sampled at random from the existing data distribution. This calculation is done by binning the most promising materials. In parenthasees we provide the results from modelling the data distributions as Gaussian and integrating to find the cumulative probability; we see alignment generally within a factor of two

in these numbers. Significant likelihoods are in bold. The bromides, chlorides, and iodides are the best candidates for a fast conducting, >4V window width structure. The chlorides and bromides are also particularly likely to exhibit fast ion conduction and possess an oxidation potential above 4 V. If ionic conductivity is unimportant and only high anodic stability is desired, the flourides are the most promising candidates. The nitrides and phosphides are the most promising class of materials for fast ion conductors that are stable against lithium metal. If ionic conductivity is not a requirement, the bromides are promising as well. No families have significant likelihoods of yielding materials that are fast conducting, stable against lithium metal, and have a >4V oxidation potential. This suggests that it is much more likely to realize a >4V solid-state Li metal battery with two electrolytes rather than one ideal material; the Gaussian integration method suggests the likelihood of success of a double-electrolyte architecture is approximately two magnitudes greater.

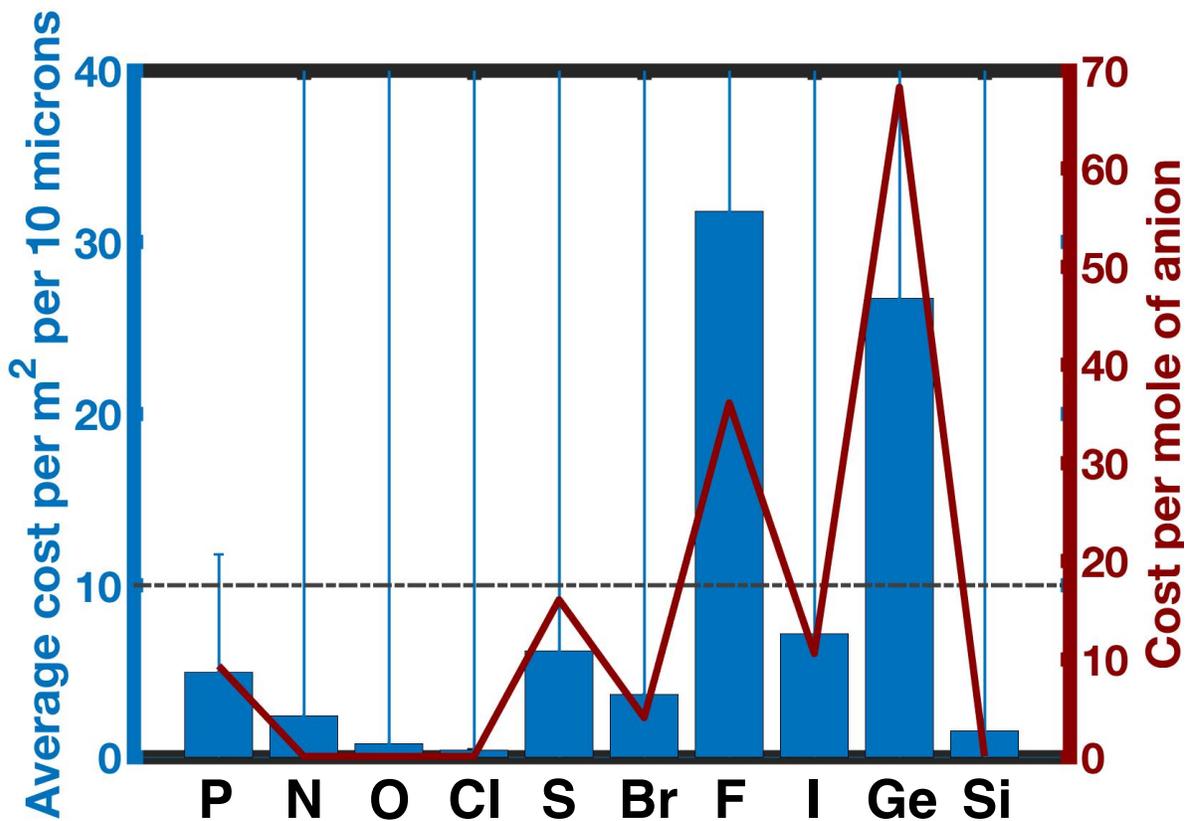

**Figure 3:** *Estimated raw materials costs by family.* On the left axis we provide the median estimated raw materials costs per m$^2$ per 10 micron thickness of stable lithium-containing solids sorted by the lattice anion. At this thickness, all families except the fluorides and germanides have an average cost under the DOE target of $10/m$^2$, denoted by the dotted grey line. On the right axis is the cost per mole of the anion. The standard deviation of the data in each family is enormous, suggesting that cost must be considered on a case-by-case basis.

|  | Superionic likelihood | Electrochemical stability window width | Oxidation potential | Reduction potential | Materials cost | Anion electronegativity |
| --- | --- | --- | --- | --- | --- | --- |
| Superionic likelihood | 1.0 | -0.10 | -0.49 | -0.28 | 0.08 | -0.71 |
| Electrochemical stability window width | -0.10 | 1.0 | 0.41 | -0.26 | -0.02 | 0.19 |
| Oxidation potential | -0.49 | 0.41 | 1.0 | 0.50 | 0.04 | 0.72 |
| Reduction potential | -0.28 | -0.26 | 0.50 | 1.0 | 0.09 | 0.49 |
| Materials cost | 0.08 | -0.02 | 0.04 | 0.09 | 1.0 | 0.03 |
| Anion electronegativity | -0.71 | 0.19 | 0.72 | 0.49 | 0.03 | 1.0 |

**Table 2:** *Pearson correlation coefficients for the observed correlations between important solid electrolyte properties.* Correlations between predicted performance metrics in hundreds to thousands of candidate solid Li-ion electrolyte materials are quantified here. We note a negative correlation between superionic likelihood and the three different indicators of electrochemical stability. We also note that the electronegativity of the anion in the lattice exhibits substantive correlations with most properties except materials cost. Materials cost shows low correlations with all properties.

# SUPPLEMENTAL INFORMATION
# Quantifying the search for solid Li-ion electrolyte materials by anion: a data-driven perspective

*S1. Data distributions in performance space*

In order to quantify the likelihood of outlier discoveries in each material family, we employ two methods. In the first, we count the fraction of all materials in each family that possess predicted performance values beyond the minimum required metrics ("binning"). The second method entails assigning a Gaussian probability distribution for each family and computing the cumulative probability of the distribution that lies beyond the minimum required performance metrics ("integrating"). The integration approach assumes that the predictive metrics we employ here (data-driven ionic conductivity classifier and DFT-based electrochemical stability calculations) are sufficiently predictive as to elucidate the general expected performance distribution of each family on the whole, even if some error exists within individual predictions. We find good agreement between both the counting method and the Gaussian distribution method, as provided in Table 1 of the main text, suggesting these calculations are fairly robust with respect to the two different methods.

The accuracy of the Gaussian distribution method relies on the data being approximately Gaussian distributed. In order to assess the Gaussian character of the data, we provide in Figs. S1-10 normalized histograms of all four performance metrics for each family: band gap, oxidation potential, reduction potential, and superionic likelihood. For each distribution we perform the Shapiro-Wilk test for normality[1] and provide the resulting *p*-values in the legend. We note that the Shapiro-Wilk test is very sensitive to outliers, which brings the *p*-value down in several cases despite a Gaussian-looking distribution. A minimum *p*-value of 0.05 for having confidence in the Gaussian character of the data is commonly used.

The Gaussian distribution for each family in two-dimensional Gaussian performance space is defined by mean $\mu$ and covariance $\Sigma$, which we denote $\mathcal{N}(\mu, \Sigma)$. We calculate $P_{\text{outlier}}$ by integrating the Gaussian over the upper-right quadrant of the graph. For the case of superionic behavior and a wide electrochemical stability window width $V_w$ (Section *II,i*), we assume $V_w \approx E_{\text{gap}}$ and integrate the section of the distributions over the desired minimum window width $v_{w,\text{min}}$:

$$P_{\text{outlier}}(V_w \geq V_{w,\text{min}}) = \frac{\int_{0.5}^{1} dP_{\text{superionic}} \int_{V_{w,\text{min}}}^{\infty} dV \, \mathcal{N}(\mu, \Sigma)}{\int_{0}^{1} dP_{\text{superionic}} \int_{-\infty}^{\infty} dV \, \mathcal{N}(\mu, \Sigma)}$$

The numerator represents the Gaussian integrated over the upper-right quadrant of the graph, and the denominator represents the Gaussian integrated over the entire graph as a normalization factor. The $P_{\text{outlier}}$ values corresponding to a minimum stability window width $V_{w,\text{min}} = 4$ V represent the likelihood that a new material from that family will have an electrochemical stability window of 4 V or wider. The likelihoods associated with $V_{w,\text{min}} = 4$ V are provided in Table 1 along with the values computed from the binning method. We see good agreement between these methods.

In Section *II,ii*, we compute the likelihood of discovering a material with superionic conduction and an oxidation potential $V_{\text{ox}}$ above some minimum value $V_{\text{ox,min}}$:

$$P_{\text{outlier}}(V_{\text{ox}} \geq V_{\text{ox,min}}) = \frac{\int_{0.5}^{1} dP_{\text{superionic}} \int_{V_{\text{ox,min}}}^{\infty} dV \, \mathcal{N}(\mu, \Sigma)}{\int_{0}^{1} dP_{\text{superionic}} \int_{-\infty}^{\infty} dV \, \mathcal{N}(\mu, \Sigma)}$$

The likelihoods corresponding to $V_{ox,min}$ = 4V are provided in Table 1.

In section *III,ii*, we compute the likelihood of discovering a material with superionic conduction and a reduction potential $V_{red}$ below some maximum value $V_{red,max}$:

$$P_{\text{outlier}}(V_{\text{red}} \leq V_{\text{red,max}}) = \frac{\int_{0.5}^{1} dP_{\text{superionic}} \int_{-\infty}^{V_{\text{red}}} dV \, \mathcal{N}(\mu, \Sigma)}{\int_{0}^{1} dP_{\text{superionic}} \int_{-\infty}^{\infty} dV \, \mathcal{N}(\mu, \Sigma)}$$

The likelihoods corresponding to $V_{red,max}$ = 0V are provided in Table 1.

In section *II,iv*, we compute the likelihood of discovering an outlier material with superionic conduction that is stable against reduction at Li metal ($V_{red,max}$ = 0 V) and stable against oxidation to at least 4 V ($V_{ox,min}$ = 4 V). Zero of the candidate materials we study here satisfy this criteria, so we cannot employ the binning approach. Quantifying the likelihood of discovering such a "wonder material" requires extrapolation of the assumed performance distributions, which the integration method allows. Treating the oxidation potential as independent of the reduction potential, the intersection of these two requirements requires a product of probabilities:

$$P_{\text{discovery}}(V_{\text{red}} \leq 0, V_{\text{ox}} \geq V_{\text{ox,min}}) = P_{\text{discovery}}(V_{\text{red}} \leq 0) \times P_{\text{discovery}}(V_{\text{ox}} \geq V_{\text{ox,min}})$$

The results of this calculation are provided in Fig. 2(g) and Table 1.

**Figure S1. Distributions of performance data by family.** For each of the ten families considered here, we visualize the (normalized) distributions of the band gap (electrochemical stability window width), oxidation potential, reduction potential, and superionic likelihood. For each distribution, we use the Shapiro-Wilks test to compute the likelihood of normality; the corresponding *p*-values are provided in the legends. The data looks approximately Gaussian in several cases, although the occasional presence of outliers can result in low *p*-values.

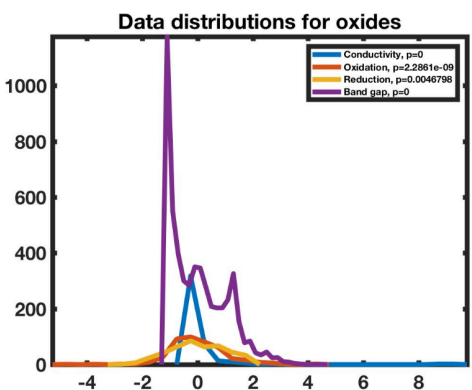
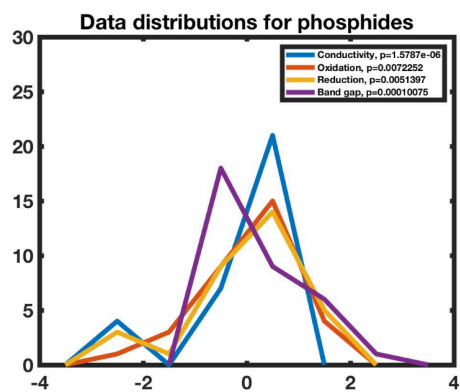
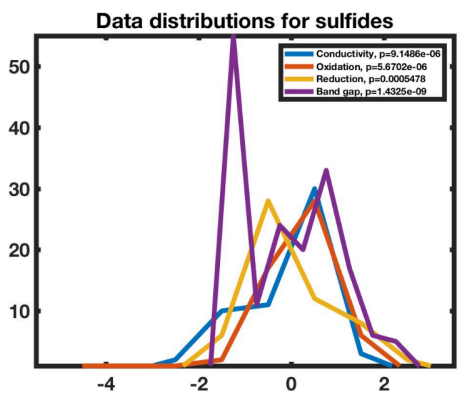
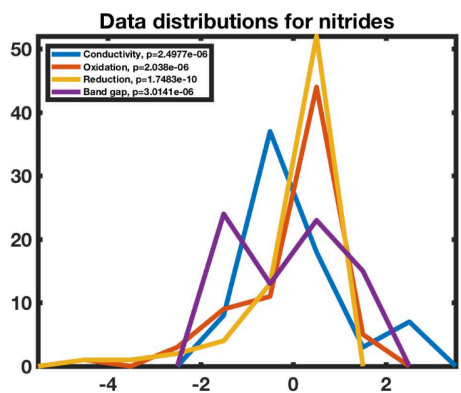

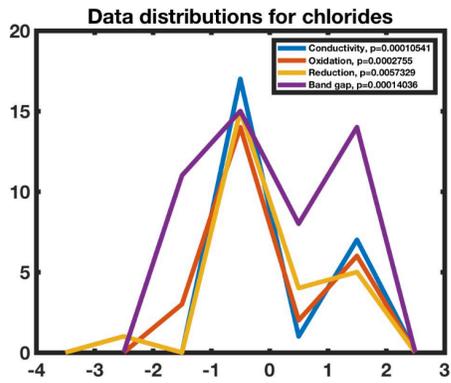
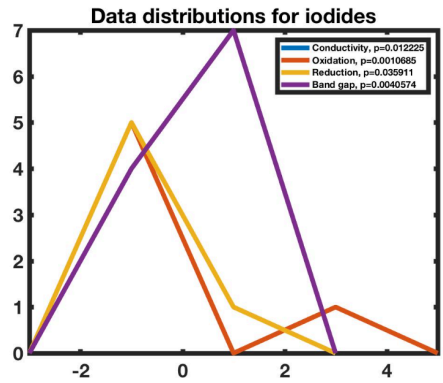
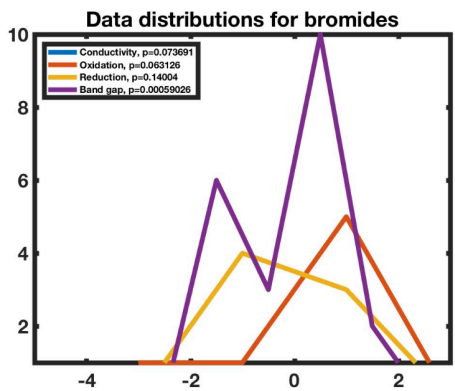
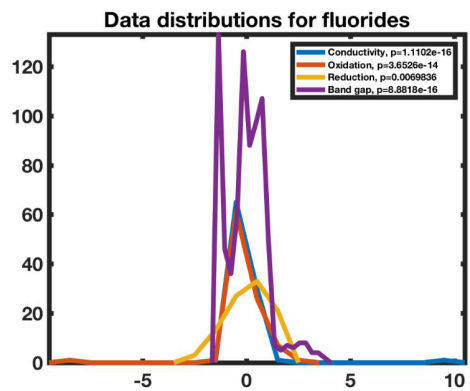
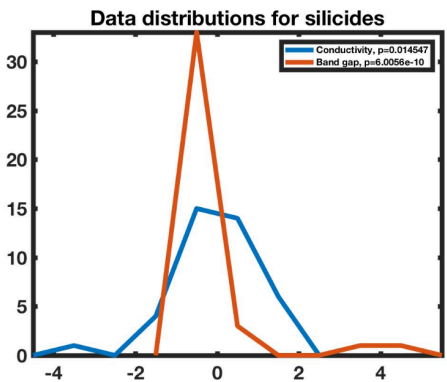
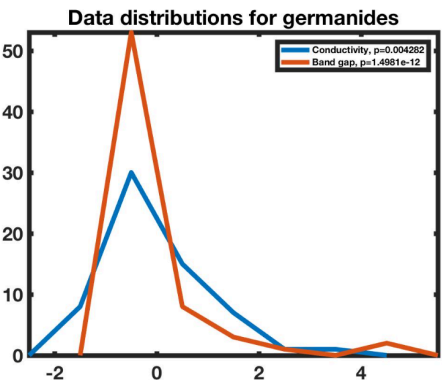

**Figure S2:** *Outlier discovery likelihood by anion family.* We use the Gaussian method to quantify outlier discovery likelihood for each of the four demands in Section II. In **(a)** we plot the likelihood of discovering an outlier material from each family that has an electrochemical stability window width exceeding the value represented on the *x*-axis. It can be seen that the fast conducting materials with window widths beyond 6V are rare, according to the structures and data available in the Materials Project database. In **(b)** we plot the likelihood of discovering an outlier material with fast ionic conductivity and a window width larger than the *x*-axis value. In **(c)** and **(d)** we perform the same calculation but replace the stability window width distribution with the oxidation potential distribution. In **(e)** we plot the likelihood of discovering a material in each family with a reduction potential below the value on the *x*-axis; *x*=0 corresponds to stability against Li metal. In **(f)** we add the ionic conductivity constraint. In **(g)** we plot the likelihood of discovering a material with stability against Li metal and an oxidation potential above the value on the x-axis; in **(h)** we add in the ionic conductivity constraint. In all plots, the error bars correspond to the standard deviations in likelihood values calculated upon removing 10% of the data at random ten times and recalculating the likelihoods.

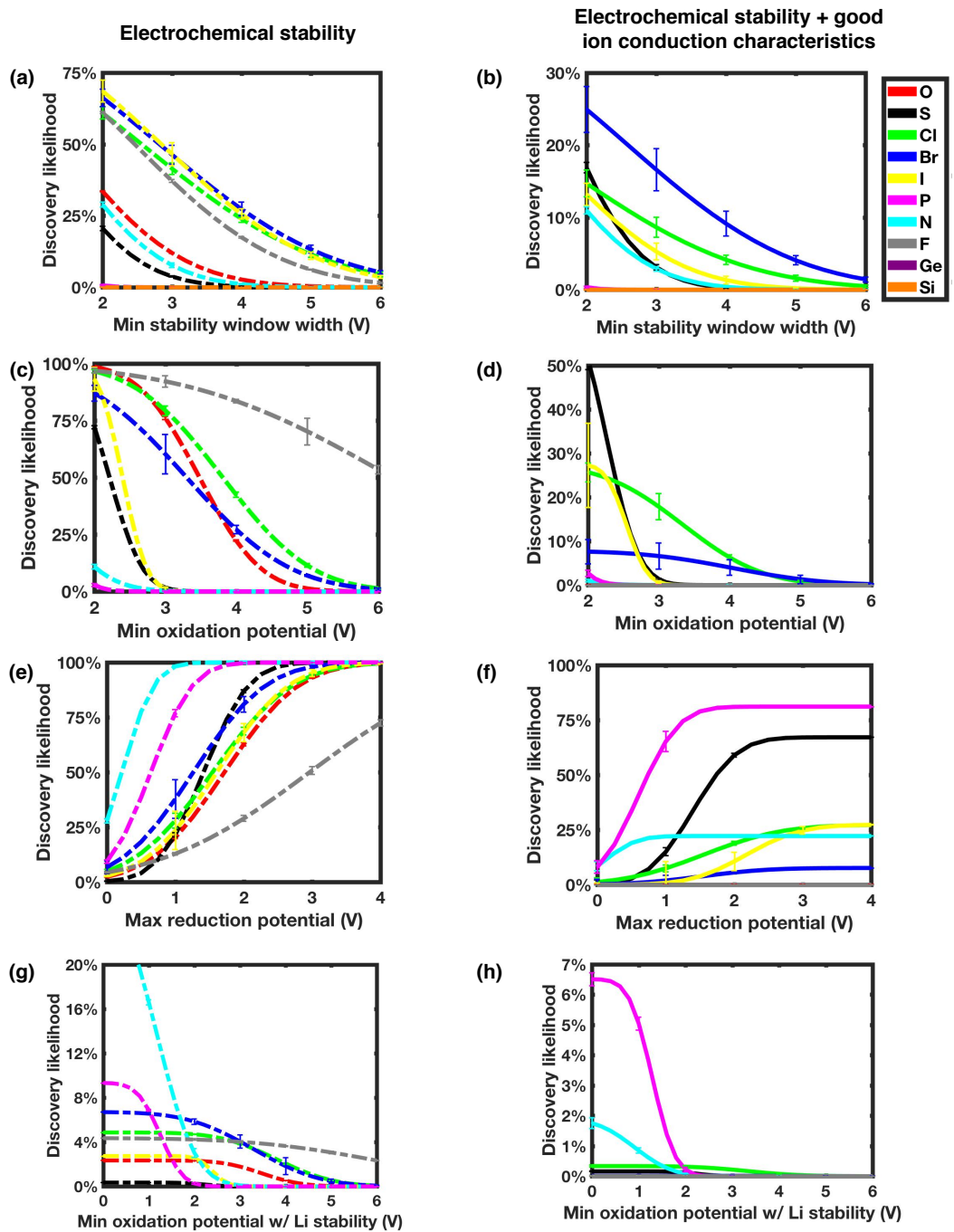